\documentclass[twocolumn]{aastex631}

\usepackage{booktabs}
\usepackage{bm}
\usepackage{threeparttable}
\usepackage{tablefootnote}
\usepackage{natbib}

\makeatletter
\renewcommand{\tablecomments}[1]{%
  \par\vspace{2pt}%
  \begin{flushleft}%
  {\normalsize Note. #1}%
  \end{flushleft}%
}
\makeatother

\begin{document}

\title{Kepler-1624b Has No Significant Transit Timing Variations}

\author[0009-0006-0871-1618]{Haedam Im}
\affiliation{Department of Physics, Massachusetts Institute of Technology, Cambridge, MA 02139, USA
}
\email{haedamim@mit.edu}

\author[0000-0001-9518-9691]
{Morgan Saidel}
\affiliation{Division of Geological and Planetary Sciences, California Institute of Technology, Pasadena, CA 91125, USA}

\author[0000-0002-5375-4725]
{Heather A. Knutson}
\affiliation{Division of Geological and Planetary Sciences, California Institute of Technology, Pasadena, CA 91125, USA}

\author[0000-0002-0371-1647]{Michael Greklek-McKeon}
\affiliation{Division of Geological and Planetary Sciences, California Institute of Technology, Pasadena, CA 91125, USA}

\author[0000-0003-2527-1475]{Shreyas Vissapragada}
\affiliation{Carnegie Science Observatories, 813 Santa Barbara Street, Pasadena, CA 91101, USA}

\author[0000-0001-6588-9574]{Karen A.\ Collins}
\affiliation{Center for Astrophysics \textbar \ Harvard \& Smithsonian, 60 Garden Street, Cambridge, MA 02138, USA}

\author[0000-0002-4909-5763]{Akihiko Fukui}
\affiliation{Komaba Institute for Science, The University of Tokyo, 3-8-1 Komaba, Meguro, Tokyo 153-8902, Japan}
\affiliation{Instituto de Astrofisica de Canarias (IAC), 38205 La Laguna, Tenerife, Spain}

\author[0000-0001-8511-2981]{Norio Narita}
\affiliation{Komaba Institute for Science, The University of Tokyo, 3-8-1 Komaba, Meguro, Tokyo 153-8902, Japan}
\affiliation{Astrobiology Center, 2-21-1 Osawa, Mitaka, Tokyo 181-8588, Japan}
\affiliation{Instituto de Astrofisica de Canarias (IAC), 38205 La Laguna, Tenerife, Spain}

\author[0000-0003-0062-1168]{Kimberly Paragas}
\affiliation{Division of Geological and Planetary Sciences, California Institute of Technology, Pasadena, CA 91125, USA}

\author[0000-0001-8227-1020]{Richard P. Schwarz}
\affiliation{Center for Astrophysics \textbar \ Harvard \& Smithsonian, 60 Garden Street, Cambridge, MA 02138, USA}

\author[0000-0002-1836-3120]{Avi Shporer}
\affiliation{Department of Physics and Kavli Institute for Astrophysics and Space Research, Massachusetts Institute of Technology, Cambridge, MA 02139, USA}

\author{Gregor Srdoc}
\affil{Kotizarovci Observatory, Sarsoni 90, 51216 Viskovo, Croatia}

\begin{abstract}

It is relatively rare for gas giant planets to have resonant or near-resonant companions, but these systems are particularly useful for constraining planet formation and migration models. In this study, we examine Kepler-1624b, a sub-Saturn orbiting an M dwarf that was previously found to exhibit transit timing variations with an amplitude of approximately 2 minutes, suggesting the presence of a nearby non-transiting companion. We reanalyze the transits from archival Kepler data and extend the TTV baseline by 11 years by combining TESS data with three new ground-based transit observations from Palomar and Las Cumbres Observatories. We jointly fit these datasets and find that the TTV amplitude is significantly weaker in our updated analysis. We calculate the Bayes factor for a one-planet versus two-planet model and find that the one-planet model is preferred.  Our results highlight the need for careful analysis of systems with relatively low amplitude TTV signals that are identified in large automated catalogs.
\end{abstract}


\section{Introduction} \label{sec:intro}

Transit and radial velocity surveys have revealed that M dwarfs frequently host compact systems of multiple small ($<3~R_\oplus$) planets \citep[e.g.,][]{DressingCharbonneau2015, MuldersPascucci2015, HsuFord2020, MignonDelfosse2025}, and that the occurrence rate of small planets around M stars is higher than for Sun-like stars \citep{DressingCharbonneau2015, RibasReiners2023, MignonDelfosse2025}. \citet{ChachanLee2023} proposed that this can be explained by the higher pebble accretion efficiency in the inner disks of M dwarfs versus Sun-like stars.  In contrast, the occurrence rate of gas giants around M dwarfs is substantially lower than that of Sun-like stars \citep{BonfilsDelfosse2013, MontetCrepp2014,  BryantBayliss2023, PassWinters2023, MignonDelfosse2025}. This is likely due to a combination of factors, including: (1) lower-mass stars tend to have less massive protoplanetary disks \citep{AndrewsRosenfeld2013} and thus most have insufficient material to form large planetary cores \citep[e.g.,][]{ChachanLee2023, KanodiaMahadevan2023, DelamerKanodia2024}, (2) the longer orbital timescales of M dwarfs slow down core formation \citep{LaughlinBodenheimer2004}, and (3) M dwarf disks dissipate faster due to intense UV radiation \citep{AdamsHollenbach2004}. These factors make it difficult for M dwarf systems to form the massive cores required for runaway gas accretion before the disk dissipates \citep{BurnSchlecker2021, KanodiaLibby-Roberts2022}. 

\begin{figure}[t!]
    \hspace*{-0.6cm}
    \centering \includegraphics[width=1.1\columnwidth]{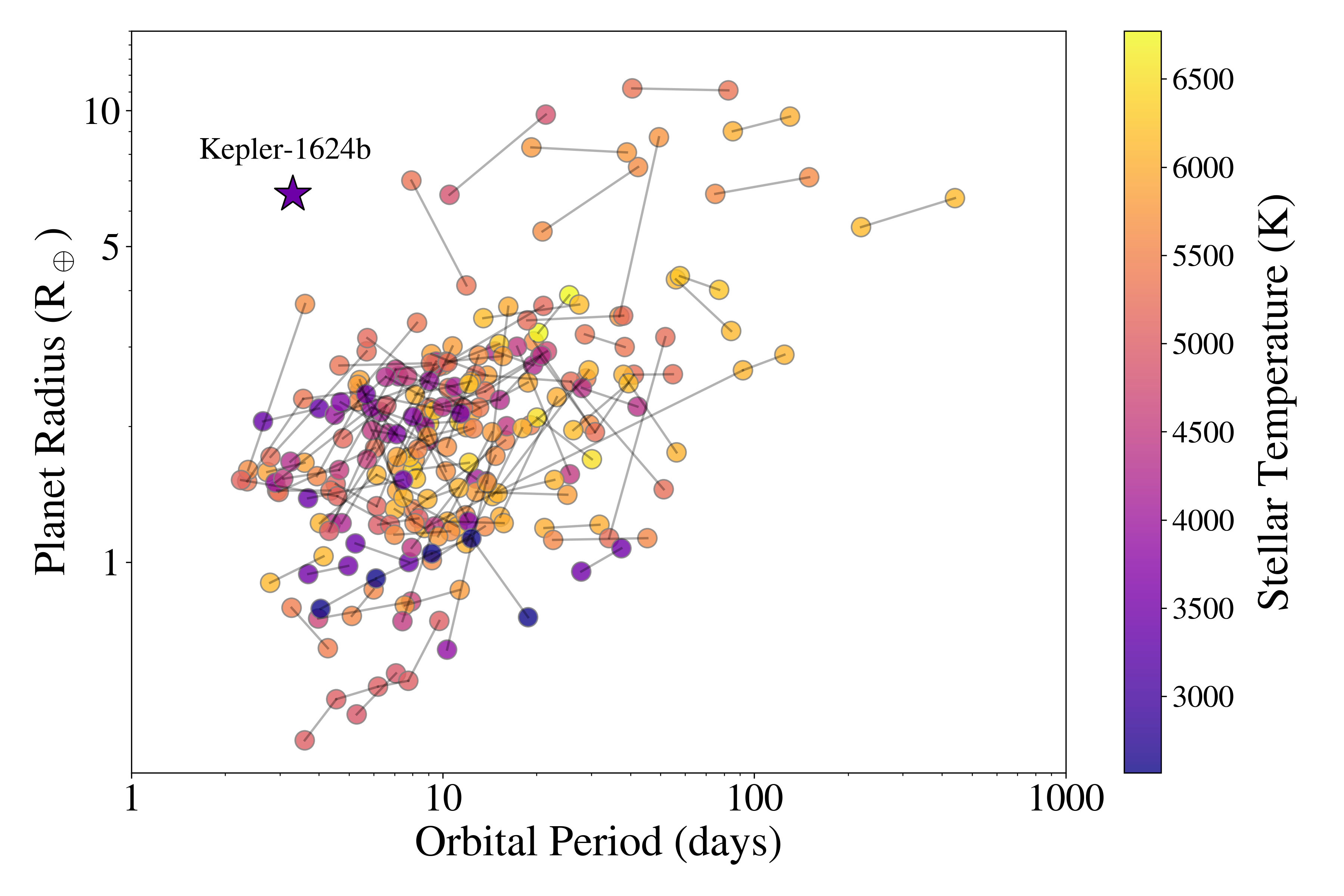} 
    \caption{Orbital period versus planet radius for all planet pairs with period ratios within 2\% of a first-order resonance.  The colors of the points indicate the effective temperature of the host star, and each planet pair is connected by a solid black line.  The location of Kepler-1624b in this space is marked by a star. 
    This plot was generated using data from the NASA Exoplanet Archive \citep{Akeson2013}. }
    \label{fig:Introplot}
\end{figure}

For stars of all masses, models of giant planet formation suggest that these planets are most easily formed at orbital separations greater than a few au \citep[e.g.,][]{Bodenheimer2000,IdaLin2004,IdaLin2008,Morbidelli2015,Chachan2021}.  This means that most giant planets on close-in orbits likely migrated inward from a more distant formation location.  However, we currently have very few observational constraints on migration mechanisms for gas giants orbiting M dwarfs.  One way to obtain these constraints is to look for the presence of companions located near orbital resonances \citep[e.g.,][]{Huang2016,Wu2023}.  The presence of these companions is a hallmark of disk migration, which makes it more likely for these small planets to be captured into orbital resonances with the gas giant \citep[e.g.,][]{CharalambousTeyssandier2022,Dai2024,Keller2025}. 
However, when we plot the sample of all confirmed planets with nearby companions within 2\% of a first-order orbital resonance, the few gas giants ($R>4~R_\oplus$) in the sample all orbit G and K stars (Figure \ref{fig:Introplot}).

It is possible that these companions exist, but have not yet been identified in the literature.  Transit timing variations (TTVs)---deviations from the periodic transit times expected for a planet on a Keplerian orbit---provide a convenient method to search for these companions using existing survey data \citep[e.g.,][]{Wu2023}. TTVs result from dynamical interactions between planets and are particularly pronounced for planet pairs near first-order mean motion resonances \citep{DeckAgol2016, AgolFabrycky2018}. \citet{HolczerMazeh2016} compiled a catalog based on the Kepler survey that identified systems with statistically significant TTVs. 15 of the systems identified in this catalog had M dwarf hosts ($T_{\text{eff}} < 3900 \text{ K}$), but only one (Kepler-1624, also known as KOI-4928) contained a transiting gas giant planet ($R_p>4~R_\oplus$) with TTVs.
\citet{HolczerMazeh2016} reported a relatively small TTV amplitude of $2.120\pm0.680$ minutes for Kepler-1624b, and \citet{GajdosVanko2019} also reported a similar amplitude of $1.9\pm 0.2$ minutes. However, a more recent reanalysis of this system by \citet{KayeAigrain2025} incorporating new TESS observations did not find evidence for statistically significant TTVs in this system. 
In this paper, we investigate the TTVs of Kepler-1624b \citep{MortonBryson2016}, a Neptune-sized exoplanet ($R_{\text{p}} = 6.59_{-0.17}^{+0.17}~R_\oplus$, $P = 3.2903045 \pm 4.6 \times 10^{-6}$ days) orbiting an M dwarf star.\footnote{We quote transit shape and ephemeris parameters from \cite{MortonBryson2016} but recalculate $R_{\text{p}}$ using stellar parameters rederived in this work.}
We reanalyze all available Kepler and TESS data along with three new high-precision observations, each obtained by the Wide-field InfraRed Camera (WIRC) on the 200 inch Hale Telescope at Palomar Observatory, the SINISTRO camera on the 1~m telescope at McDonald Observatory, and MuSCAT3 on the 2~m Faulkes Telescope North (FTN), respectively. In \S\ref{sec: observe}, we describe the details of each observation and our fits to obtain transit midtimes. In \S\ref{Dynamical Modeling}, we detail the dynamical model used to interpret our TTVs. In \S\ref{sec: discus and conc}, we present the results of our analysis and assess the statistical significance of the TTV signal. Lastly, in \S\ref{sec: conclusion}, we discuss the broader implications of this study. 

\begin{figure*}[t]
    \centering
    \includegraphics[width=\textwidth]{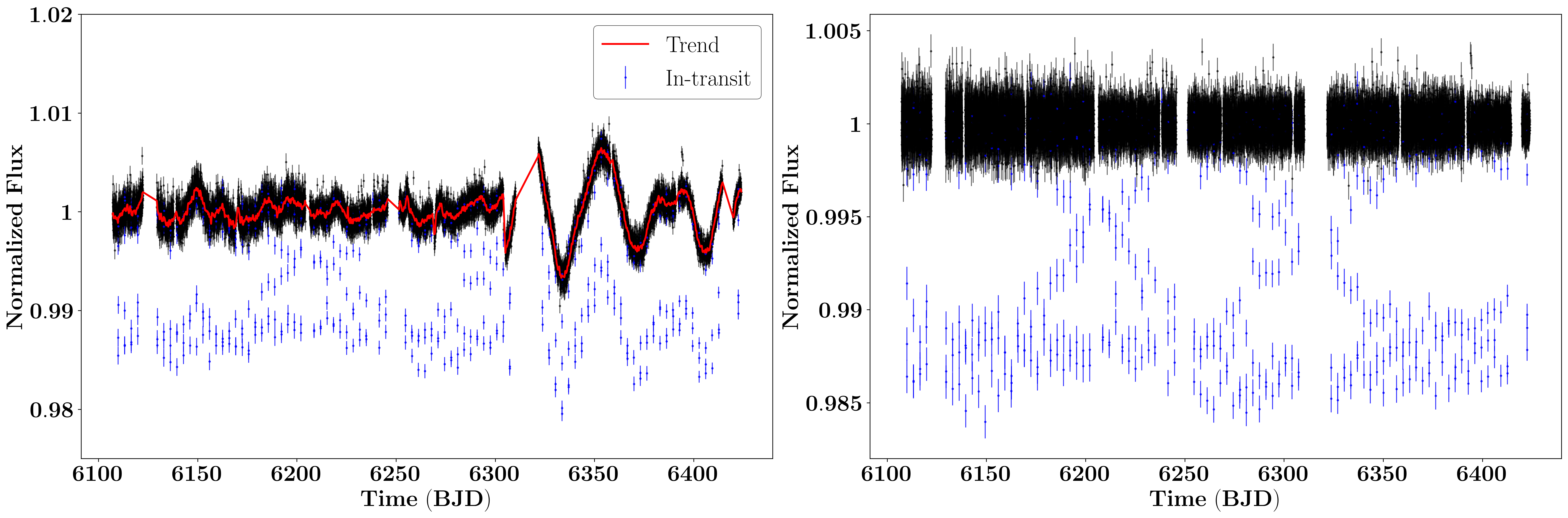}
    \caption{Kepler light curve before detrending (left) and after detrending (right). The red line shows the Gaussian kernel convolution trend described in \S\ref{Kepler}. Blue points indicate masked data ($\pm 0.8$ transit durations around each transit midtime).}
    \label{detrending_plot}
\end{figure*}

\begin{figure*}[htbp]
    \centering
    \includegraphics[width=1\textwidth]{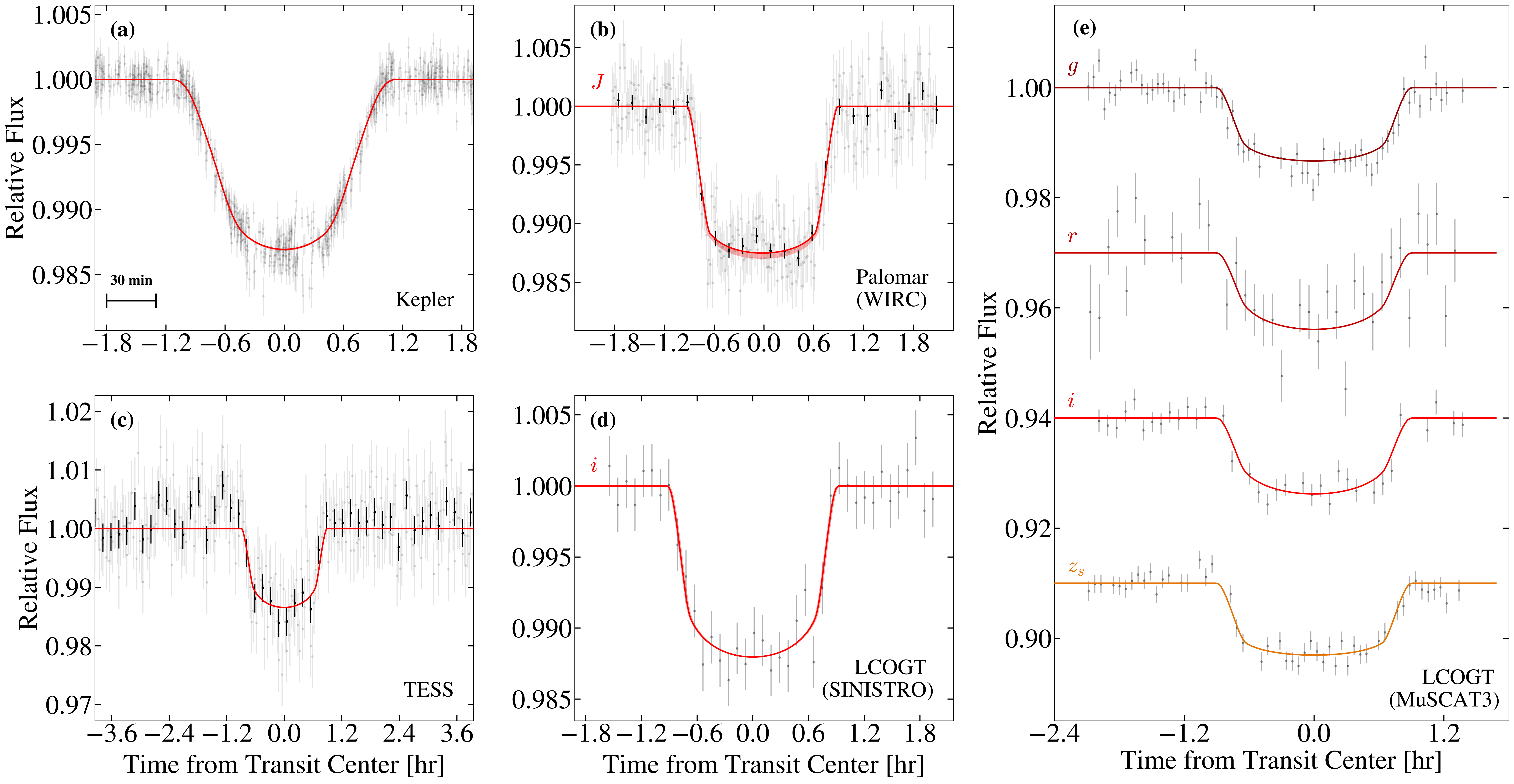} \\[0.3cm]
    \caption{Transit light curves for Kepler-1624b from multiple instruments. (a) Kepler, (b) Palomar/WIRC, (c) TESS, (d) LCOGT/SINISTRO $i$ band, and (e) LCOGT/MuSCAT3 multiband. For Kepler, the light curves of individual transits are phased up using the individual best-fit transit times.  For TESS, we phase individual transits using transit times from the best-fit one-planet model.  For (a), the gray points show unbinned phased data, where each point corresponds to a 30-minute integration. For (b), unbinned photometry is shown as gray circles, and 10-minute binned photometry is shown as black circles. For (c), we use 10-minute bins for the black points and 2-minute bins for the gray points. For (d), we show the unbinned data as gray points. 
    For (a)--(d), the best-fit transit light curves are overplotted in red, with the 1$\sigma$ confidence range overplotted as light red shading. For (e), transits in  $g$, $r$, $i$, and $z_s$ bands are plotted with different colors with an arbitrary vertical offset, and the gray points represent the observed data, while the colored lines show the best-fit transit model for each band with a 1$\sigma$ confidence range overplotted with the corresponding color of shading. For some of the panels, the 1$\sigma$ confidence shading is not visually noticeable due to very tight confidence ranges. } 
    \label{fig:transit panels}
\end{figure*}

\section{Observations and Data Reduction} \label{sec: observe}
\subsection{Updated Stellar Parameters}

We derived updated constraints on the stellar parameters for Kepler-1624 using $T_{\mathrm{eff}} = 3736 \pm 60$~K and [Fe/H] $= 0.28 \pm 0.08$ dex from \cite{BirkyHogg2020} based on fits to APOGEE spectroscopy \citep{MajewskiSchiavon2017}, Gaia parallax $\varpi = 3.933 \pm 0.027$ mas \citep{GaiaCollaborationVallenari2023}, and Two Micron All Sky Survey (2MASS) infrared photometry \citep{CutriSkrutskie2003}. We combined this information to derive updated stellar parameters using the \texttt{isoclassify} package, which matches observational data to pre-computed stellar evolution MESA models using a grid-based approach \citep{Huber2017}.  This yielded an effective temperature $T_{\mathrm{eff}} = 3768^{+42}_{-44}$~K, surface gravity $\log g = 4.685^{+0.011}_{-0.012}$~cm~s$^{-2}$, metallicity [Fe/H] $= 0.269^{+0.066}_{-0.068}$ dex, stellar radius $R_* = 0.552^{+0.014}_{-0.013}~R_\odot$, and stellar mass $M_* = 0.541^{+0.012}_{-0.012}~M_\odot$. These values are in good agreement with the ones reported in \cite{StassunOelkers2019}, which used a similar approach but without the benefit of a spectroscopic constraint on the stellar effective temperature and metallicity.

\subsection{Kepler \label{Kepler}}

The Kepler-1624 system was observed during Kepler Quarters 14 through 17, all in long-cadence (LC) mode \citep{keplerq14, keplerq15, keplerq16, keplerq17}. We obtained the Pre-search Data Conditioning Simple Aperture Photometry (PDCSAP) light curves from the Mikulski Archive for Space Telescopes. These data were preprocessed using the PDC cotrending algorithm \citep{StumpeSmith2012, SmithStumpe2012, StumpeSmith2014} by the Kepler team to remove systematic trends. 

We fit the PDCSAP data from Kepler Data Release 25 to extract transit midtime measurements. First, we detrended the data using a Gaussian kernel convolution and remove any stellar variability following the approach described in \citet{Greklek-McKeonKnutson2023}. We masked $\pm 0.8$ transit durations around each transit midtime reported in \citet{HolczerMazeh2016} to prevent in-transit data from influencing the trend model. 
We found gaps where consecutive data points in the times series had a separation larger than 0.5 days. Using these gaps, we divided the data into multiple segments. For each segment, we applied a separate Gaussian kernel convolution to estimate long-term trends and set the kernel width to 1.5 times the transit duration.
 Convolutions done near the boundaries of time series can introduce artificial features if there are insufficient data points for the kernel to operate on. Rather than using simple padding or truncation, we utilized a reflective method. At both ends of our segments, we selected a quarter of the data and created a temporal reflection of reversed data that extended the boundaries of the segment. The temporal reflection was then adjusted to an appropriate baseline based on the mean flux level of the quarters. 
We then divided the original flux values and their corresponding uncertainties by the trend model to produce a detrended light curve. Figure \ref{detrending_plot} displays the detrended light curve.

After detrending, we extracted individual transit events by selecting data within $\pm 3$ transit durations around each transit midtime reported by \citet{HolczerMazeh2016}, excluding any transits with fewer than five data points, or 2.5 hours, in the selected window. This resulted in a total of 84 individual transit observations. We used the transit midtimes reported in \citet{HolczerMazeh2016} to phase up the transits to make a stacked Kepler transit light curve. We fit this phased transit light curve with the \texttt{BATMAN} package \citep{Kreidberg2015}, which parameterizes the transit shape using the ratio of the semi-major axis to the stellar radius $a/R_*$, the orbital inclination $i$, the planet-to-star radius ratio $R_p/R_*$, and the quadratic limb darkening parameters. We used 20 evenly spaced samples over the exposure time to supersample the average value of the light curve over the long-cadence integration time of 30 minutes. 

We set Gaussian priors on the quadratic limb darkening coefficients centered on the values calculated by \texttt{ldtk} \citep{ParviainenAigrain2015} with uncertainties inflated by a factor of 10. For $a/R_*$, we used a Gaussian prior of $13.7\pm0.6$, which was calculated from Kepler's third law using the stellar mass and radius derived from \texttt{isoclassify}. We adopted a uniform prior from $60^\circ$ to $90^\circ$ for orbital inclination $i$ and from 0.0 to 0.2 for planet-to-star radius ratio $R_p/R_*$. We also fit the photometric uncertainty of the Kepler data with a uniform prior ranging from 0 to 1000 ppm. We explored the parameter space using a Markov chain Monte Carlo algorithm implemented by \texttt{emcee} \citep{Foreman-MackeyConley2013} with 50 walkers, 3000 burn-in steps, and 10,000 steps. We confirmed that the total chain length exceeded 50 autocorrelation lengths for all parameters.

We then fit for the individual transit midtimes while keeping all transit parameters fixed to the best-fit results from the phased transit. We used a uniform prior of $\pm$ 15 minutes from the transit midtime reported from \citet{HolczerMazeh2016} for each transit midtime. We then phased up the transits again using our newly retrieved midtimes, solved for the best-fit transit parameters, and repeated our fits using the updated transit shape parameters to determine the final set of individual transit midtimes. 

\begin{deluxetable}{lcc}[t]
\tablecaption{Transit Shape Parameters from the Joint Fit to the Phased Kepler and Palomar/WIRC Light Curves}
\tablewidth{0pt}
\tablehead{
\colhead{Parameter} & \colhead{Priors} & \colhead{Best-fit Values}}

\startdata
$P$ (days)$^a$ & $\mathcal{N}(3.290314, 0.00015)$ & $3.290314_{-0.000015}^{+0.000015}$ \\
$t_{0}$$^a$ & -- & $6290.90198_{-0.00011}^{+0.00011}$ \\
$b$ & $\mathcal{U}(0, 1.2)$ & $0.553_{-0.083}^{+0.060}$ \\
$a/R_*$ & $\mathcal{U}(11.5, 15.5)$ & $13.53_{-0.67}^{+0.78}$ \\
${R_p/R_*}_{\textrm{\scriptsize WIRC}}$& $\mathcal{N}(0.10992, 0.00083)$ & $0.10939_{-0.00076}^{+0.00075}$ \\
${R_p/R_*}_{\textrm{\scriptsize Kepler}}$
 & $\mathcal{U}(0, 0.5)$ & $0.1081_{-0.0017}^{+0.0016}$ \\
$R_p~(R_{\oplus})^b$ & -- &  $6.59_{-0.17}^{+0.17}$ \\
$u_{1,\textrm{\scriptsize Kepler}}$ & $\mathcal{N}(0.510, 0.016)$ & $0.4965_{-0.0161}^{+0.0157}$ \\
$u_{2,\textrm{\scriptsize Kepler}}$ & $\mathcal{N}(0.185, 0.026)$ & $0.1684_{-0.0261}^{+0.0261}$ \\
$\sigma_{\textrm{\scriptsize WIRC}}$$^c$ & $\mathcal{U}(0.000001, 0.01)$ & $0.00160_{-0.00013}^{+0.00014}$ \\
$\sigma_{\textrm{\scriptsize Kepler}}$$^d$ & $\mathcal{U}(0.5, 5)$ & $2.240_{-0.034}^{+0.035}$ 
\enddata
\tablenotetext{a}{Period $P$ and transit midpoint $t_{0}$ are calculated from the posteriors to the one planet TTV fit including the Kepler, Palomar, TESS, and LCOGT observations as described in \S\ref{Dynamical Modeling}. $t_{0}$ is in units of BJD $-$ 2450000.}
\tablenotetext{b}{Planetary radius $R_p$ is derived using ${R_p/R_*}_{\textrm{\scriptsize WIRC}}$ and  stellar radius $R_* = 0.552^{+0.014}_{-0.013}~R_\odot$. }
\tablenotetext{c}{$\sigma_{\textrm{\scriptsize WIRC}}$ is the jitter term added in quadrature to the photometric uncertainties of the Palomar/WIRC data.}
\tablenotetext{d}{$\sigma_{\textrm{\scriptsize Kepler}}$ is the error scaling factor multiplied by the photometric uncertainties of Kepler data.}
\tablecomments{The limb darkening coefficients were fixed as $u_1 = 0.24$ and $u_2 = 0.16$ for Palomar/WIRC.}
\label{tab:Joint Fit Table}
\end{deluxetable}

\subsection{Palomar/WIRC \label{sec:Palomar}}
We observed one full transit of Kepler-1624b in $J$ band using the Wide-field InfraRed Camera (WIRC) on the 200-inch Hale Telescope at Palomar Observatory on UT 2020 August 6 \citep{WilsonEikenberry2003}. We used a beam-shaping diffuser that creates a 3\farcs0 FWHM top-hat point-spread function (PSF) across its 8\farcm7 $\times$ 8\farcm7 field of view. This custom diffuser
mitigates time-correlated noise from PSF variations, allowing WIRC to achieve a photometric precision comparable to space-based observations \citep{StefanssonMahadevan2017, VissapragadaJontof-Hutter2020}. 
The observation procedures followed those described by \citet{VissapragadaJontof-Hutter2020}. 
Science images were taken with an exposure time of 45 s and the airmass varied from 1.09 to 2.74 during the observation window. 

We used the WIRC Data Reduction Pipeline \texttt{exowirc} to analyze the WIRC data \citep{VissapragadaJontof-Hutter2020}. First, all images were flat-fielded and dark-subtracted using a median flat and a median dark frame, respectively. Then, we used the \texttt{photutils} package \citep{BradleySipocz2016} to perform differential photometry.  We considered circular apertures with radii between 5 to 20 pixels and found that a radius of 10 pixels (2\arcsec.50) yielded the lowest root mean square scatter in the normalized light curve. We utilized \texttt{DAOStarFinder} to identify and extract photometry for a sample of comparison stars, and selected the 5 stars whose light curves most closely tracked that of our target star to correct for time-dependent flux variations. We carried out an initial rough sky subtraction by calculating the median background across each image using iterative 3$\sigma$ clipping with a maximum of five iterations. We determined the location of the star in each image via iterative flux-weighted centroiding, and then estimated and subtracted the local sky background from each star using an annulus with an inner radius of 25 pixels and an outer radius of 50 pixels, applying iterative 2$\sigma$ clipping to reconstruct the mean background.

We constructed an initial systematics model consisting of a linear function of time and weights for our five comparison star light curves.  We also considered additional linear detrending vectors, including the PSF width, airmass, sky background from the median of the dithered background, and the distance moved from the initial centroid position on the detector. We chose the combination with the lowest Bayesian information criterion (BIC) value compared to the model without additional covariates.  We found that the BIC was minimized when we included the sky background along with our five comparison stars and linear function of time.

We then jointly fit the Palomar transit with the phased Kepler transit. We modeled the transit light curve using the \texttt{exoplanet} package \citep{Foreman-MackeyLuger2021}, which utilizes the Hamiltonian MCMC package \texttt{PyMC3} \citep{SalvatierWiecki2016} to map the posterior probability distribution for the model. We assumed a single global value for the orbital period $P$, impact parameter $b$, and the scaled semi-major axis $a/R_*$, while fitting for separate planet-to-star radius ratios and transit midtimes. We also fit a jitter term $\sigma_{\text{{WIRC}}}$, which is added in quadrature to the photometric uncertainties of WIRC data, and a multiplicative error scaling factor $\sigma_{\text{{Kepler}}}$ to the photometric uncertainties of Kepler data. For Kepler data, we fixed the limb darkening parameters to 
$u_1 = 0.50$ and $u_2 = 0.17$, the best-fit values from the phased Kepler fit. For Palomar/WIRC, we used $u_1 = 0.24$ and $u_2 = 0.16$ using package \texttt{ldtk} as described in \S\ref{Kepler}.

Table \ref{tab:Joint Fit Table} records the best-fit values and uncertainties for the transit shape parameters derived from this joint fit. The two panels in the top row of Figure \ref{fig:transit panels} show the phased data and best-fit transit models for the Kepler and Palomar/WIRC fits, respectively.

\subsection{Las Cumbres Observatory (LCOGT)}

We observed two full transits of Kepler-1624b using the Las Cumbres Observatory Global Telescope (LCOGT; \citealt{BrownBaliber2013}). The first transit was observed on UT 2024 October 11 with the SINISTRO camera, mounted on the 1~m telescope at McDonald Observatory. The camera has a 4096 $\times$ 4096 pixel CCD, an image scale of 0\farcs389 per pixel, and a field of view of 26\farcm5 $\times$ 26\farcm5. Observations were conducted in the SDSS $i$ band with an exposure time of 300 s. The second transit was observed on UT 2024 October 24 with MuSCAT3 \citep{NaritaFukui2020}, mounted on the 2~m FTN at the Haleakal\={a} Observatory in Maui, Hawai`i. MuSCAT3 is a multichannel imager with 2048 $\times$ 2048 pixels, an image scale of 0\farcs27 per pixel, and a 9\farcm1 × 9\farcm1 field of view. The transit was simultaneously observed in the $g$, $r$, $i$, and $z_s$ bands with exposure times of 300, 291, 174, and 202 s, respectively. Both transits were processed using the standard LCOGT \texttt{BANZAI} pipeline \citep{McCullyVolgenau2018}, and differential photometric data were extracted by \texttt{AstroImageJ} \citep{CollinsKielkopf2017}. 

Although in principle we could have added both LCOGT transits to the joint fit described in \S\ref{sec:Palomar}, we fit the LCOGT transits separately for simplicity and confirm that the joint fit including both LCOGT transits results in transit shape parameters that are within $1\sigma$ of the Palomar and Kepler joint fit. We fit the transit midtimes for LCOGT data using the \texttt{batman} package, by fixing all parameters to those listed in Table \ref{tab:Joint Fit Table} but allow the transit midtimes to vary with a wide uniform prior around the expected midtime. We used 50 walkers with 5000 steps for burn-in and 2000 steps for the final chain.

For LCOGT/SINSTRO, we used limb darkening parameters for $i$ band of $u_1 = 0.44$ and $u_2 = 0.19$, which we calculated using the \texttt{ldtk} package as described in \S\ref{Kepler}. The lower-right panel in Figure \ref{fig:transit panels} displays the best-fit light curve in red with the LCOGT/SINISTRO transit data in gray circles.

For LCOGT/MuSCAT3, we simultaneously fit the multiband photometry by having band-specific limb darkening parameters using the package \texttt{ldtk} based on the same stellar parameters as above. We used band-specific limb darkening parameters $u_1 = 0.69$ and $u_2 = 0.11$ for $g$ band,  $u_1 = 0.63$ and $u_2 = 0.11$ for $r$ band, $u_1 = 0.44$ and $u_2 = 0.19$ for $i$ band, and $u_1 = 0.35$ and $u_2 = 0.18$ for $z_s$ band. Figure \ref{fig:transit panels} presents the best-fit light curves for each band and the corresponding observed data points. 

\subsection{TESS}
In addition to the Kepler observations, we analyzed TESS \citep{RickerWinn2014} photometry from Sectors 40, 41, 54, 55, 74, 75, and 81, all in short-cadence (SC) mode \citep{tess}. We obtained the TESS data processed by SPOC pipeline version 5.0 using the \texttt{lightkurve} package \citep{LightkurveCollaborationCardoso2018}. Due to TESS's lower photometric precision, we found that individual transit midtimes derived from TESS have larger uncertainties compared to Kepler.

 For TESS data, we adopt similar detrending methods to those described in \S\ref{Kepler}. First, we masked $\pm0.8$ transit durations around the transit midtime predicted based on a linear ephemeris using the period and epoch reported from \citet{HolczerMazeh2016}. Then, we detrended the TESS data using a Gaussian convolution with a kernel width of 3.0 times the transit duration with the same reflective method described in \S\ref{Kepler}. We divided the original flux values and the photometric uncertainties by the trend to retrieve the detrended dataset. Lastly, we extracted individual transit data within $\pm 3$ transit times around each predicted transit midtime, which was directly incorporated into our photodynamical model described in \S\ref{Dynamical Modeling}.

\begin{table}[t]
\caption{Observed Transits of Kepler-1624b}
\centering
\begin{tabular}{cccc}
\hline \hline
Source & Epoch  & Midtimes  &
  Errors \\ 
  \hline 
Kepler &0 & 6109.9347 & 0.0011 \\
Kepler &1 & 6113.2258 & 0.0011 \\
Kepler &2 & 6116.5176 & 0.0010 \\
\vdots & \vdots & \vdots & \vdots \\
Palomar/WIRC &899 & 9067.92201 & 0.00039 \\
LCOGT/SINISTRO &1363 & 10594.62509 & 0.00099 \\
LCOGT/MuSCAT3 &1367 & 10607.78596 & 0.00039 \\
  \hline

\end{tabular}
\tablecomments{Transit midtimes are in units of BJD $-$ 2450000, and midtime measurement errors are in units of days. Only a portion of this table is shown here to demonstrate its form and content. A \href{https://content.cld.iop.org/journals/1538-3881/170/6/336/revision1/ajae1464t2_mrt.txt}{machine-readable} version of the full table is available in the published version of the article.}
\label{times_table}
\end{table}

\begin{figure*}[htbp]
    \centering
    \includegraphics[width=1\textwidth]{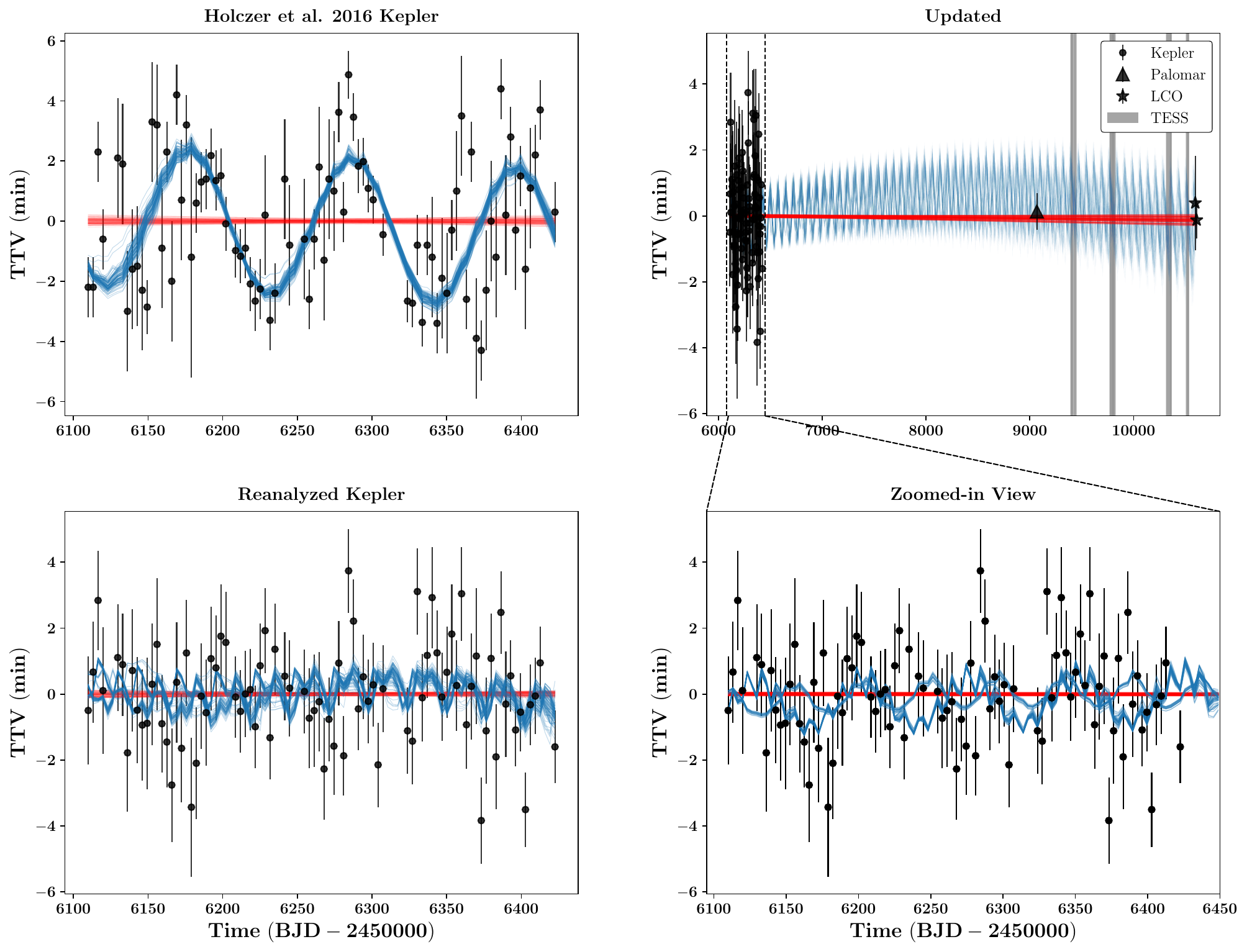} \\[0.3cm]
    \caption{Transit timing variations with dynamical models. 100 random posterior draws from the two-planet model (blue) and one-planet model (red) are plotted. The top-left panel displays the original Kepler TTVs from \citet{HolczerMazeh2016}, showing a clear sinusoidal pattern. The bottom-left panel presents our reanalysis of the Kepler data. The top-right panel displays all observed transit timings from Kepler, Palomar, and LCOGT. The time intervals where TESS transits occur are shaded in gray. The Kepler points are plotted with black circles. The Palomar point is plotted with a triangle, and the two LCOGT points are plotted with stars. The bottom-right panel presents a zoomed-in view of the Kepler data from the top-right panel.
    }
    \label{fig:ttv-comparison}
\end{figure*}

\section{Dynamical Modeling} \label{Dynamical Modeling}

We developed a hybrid photodynamical model to fit the combined information from Kepler, Palomar, TESS, and LCOGT observations.  For the Kepler, Palomar, and LCOGT data, we fit the transit midtimes, which are measured with high statistical significance in each individual transit observation.  This is not true for the TESS data, which contains many transits measured at low statistical significance.  For these data, we utilized a photodynamical model and compared the predicted transit light curves directly to the TESS photometry as described below. Although in principle we could have fit all of our light curves with this photodynamical model, this would have significantly increased the computational overheads in our model fits for a relatively minimal gain in the quality of our constraints on the planet parameters.

We used \texttt{TTVFast} \citep{DeckAgol2014} to calculate transit midtimes based on N-body simulations and calculated the corresponding likelihood for the measured Kepler, Palomar, and LCOGT transit times assuming Gaussian uncertainties on all measurements. For the TESS data, we used \texttt{batman} \citep{Kreidberg2015} to model the individual transit light curves and set the transit midtimes to the values calculated by \texttt{TTVFast}.  We fixed the transit shape parameters to the best-fit values presented in Table~\ref{tab:Joint Fit Table}. We used the \texttt{ldtk} package \citep{ParviainenAigrain2015} to calculate limb darkening parameters $u_1 = 0.40$ and $u_2 = 0.19$ for the TESS band following the process described in \S\ref{Kepler}.

We fit our dynamical model to the data using nested sampling as implemented in the \texttt{dynesty} \citep{SpeagleBarbary2018} package and fit both one- and two-planet models. This allowed us to calculate the statistical preference for a one-planet versus two-planet model using the Bayes factor. 

Each planet was defined by five parameters, including mass ($m$), orbital period ($P$), $\sqrt{e} \cos \omega$, $\sqrt{e} \sin \omega$, and a reference epoch ($t$), resulting in ten total parameters for the two-planet model and five for the one-planet model. For $m_1$ and $m_2$, we used uniform priors of 21 to 61~$M_{\oplus}$ and 3 to 40~$M_{\oplus}$, respectively. We chose these physically motivated mass ranges using the mass-radius relationship from \citet{ChenKipping2017}, to ensure that our initial fits focused on the most likely region of solution space.  We placed a Gaussian prior on $P_1$ centered on the best-fit period derived from the joint fit described in \S\ref{sec:Palomar} with an uncertainty inflated by 10 to allow for a greater range of dynamical solutions. We assumed that the second planet was most likely to be located in an exterior orbit as no additional transit signals were detected in the Kepler photometry. Models simulating an inner companion with periods smaller than that of Kepler-1624b quickly became unstable and the N-body integration terminated. Thus, we conducted an initial search with a broad period range of 4 to 13.6 days that identifies the 3:1 resonance as the best-fit solution to the observed TTVs. We then ran a second fit with a narrower uniform prior from 8.5 to 10~days around this 3:1 resonance for $P_2$. We placed broad uniform priors from $-$0.4 to 0.4 on $\sqrt{e}\sin{\omega}$ and $\sqrt{e}\cos{\omega}$ for both planets. We used a Gaussian prior for $t_1$ based on the first midtime measurement in Table \ref{times_table} with an inflated uncertainty, and set a uniform prior spanning $\pm5$ days around that first midtime value for $t_2$. For both planets, the orbital inclinations and the longitudes of the ascending nodes were fixed to 90$\degr$, respectively.
We employed a random walk sampler with 10,000 live points, which were continuously updated until a convergence of $\Delta \log \mathcal{Z}  = 0.05$ was reached, ensuring that the posterior fully explored multiple modes. We also confirmed that our Bayesian evidence $\mathcal{Z}$ was not influenced by specific prior choices by running a test with doubled prior ranges for the planets' masses and periods. 

\section{Discussion \label{sec: discus and conc}}
\subsection{Bayes Factor Comparison}
First, we quantified the initial significance of the TTV signal by running our fits on the Kepler transit times reported by \citet{HolczerMazeh2016}, where we define the Bayes factor $\Delta \log \mathcal{Z} = \log \mathcal{Z}_{\text{two-planet}} - \log \mathcal{Z}_{\text{one-planet}}$.  We found $\Delta \log \mathcal{Z}=78.2$, indicating a strong ($12.5\sigma$) preference for the two-planet model, consistent with the statistical significances reported by \citet{HolczerMazeh2016}. Figure \ref{fig:ttv-comparison} shows the resulting TTV fit with 100 random posterior draws from the two-planet model (blue) and one-planet model (red).  Next, we repeated the same fit using our updated Kepler transit times and found that the statistical significance decreased drastically, to the point where the one-planet model was preferred over the two-planet model ($\Delta \log \mathcal{Z}=-1.27$).  This preference for the one-planet model increased when we repeated our fits using our new Palomar and LCOGT transit midtimes. The inclusion of the TESS data further increased the strength of the preference for the one-planet model, with $\Delta \log \mathcal{Z}=-2.45$ ($2.73\sigma$).  We summarize these results in Table \ref{tab:model-evidence}. 

While Kepler-1624b was originally flagged as having a potential companion in \citet{HolczerMazeh2016}, our new analysis strongly favors a one-planet model instead. The $\Delta \log \mathcal{Z}$ values from the nested sampling indicate that the TTV signal diminishes as additional transits are included in the comprehensive analysis.

\begin{table}[t!]
\centering
\caption{A Summary of Bayes Factors between the Two-planet model and One-planet Model Performed on Different Combinations of Datasets}

\begin{tabular}{lcc}

\hline \hline
Tested model & $\Delta \log \mathcal{Z}$ & $\sigma$ \\
\hline
\citet{HolczerMazeh2016} Kepler & 78.2 & 12.5 \\
\citet{GajdosVanko2019} Kepler & 73.2 & 12.1 \\
Updated Kepler & $-1.27$ & 2.18 \\
Updated Kepler, Palomar, LCOGT & $-1.50$ & 2.30 \\
Updated Kepler, Palomar, TESS, LCOGT & $-2.45$ & 2.73 \\
\hline
\end{tabular}
\label{tab:model-evidence}
\tablecomments{$\Delta \log \mathcal{Z} = \log \mathcal{Z}_{\text{two-planet}} - \log \mathcal{Z}_{\text{one-planet}}$, where a positive value indicates evidence favoring the two-planet model.}
\end{table}
\subsection{Comparison to Published TTVs}

\begin{figure}
    \centering
    \includegraphics[width=1\linewidth]{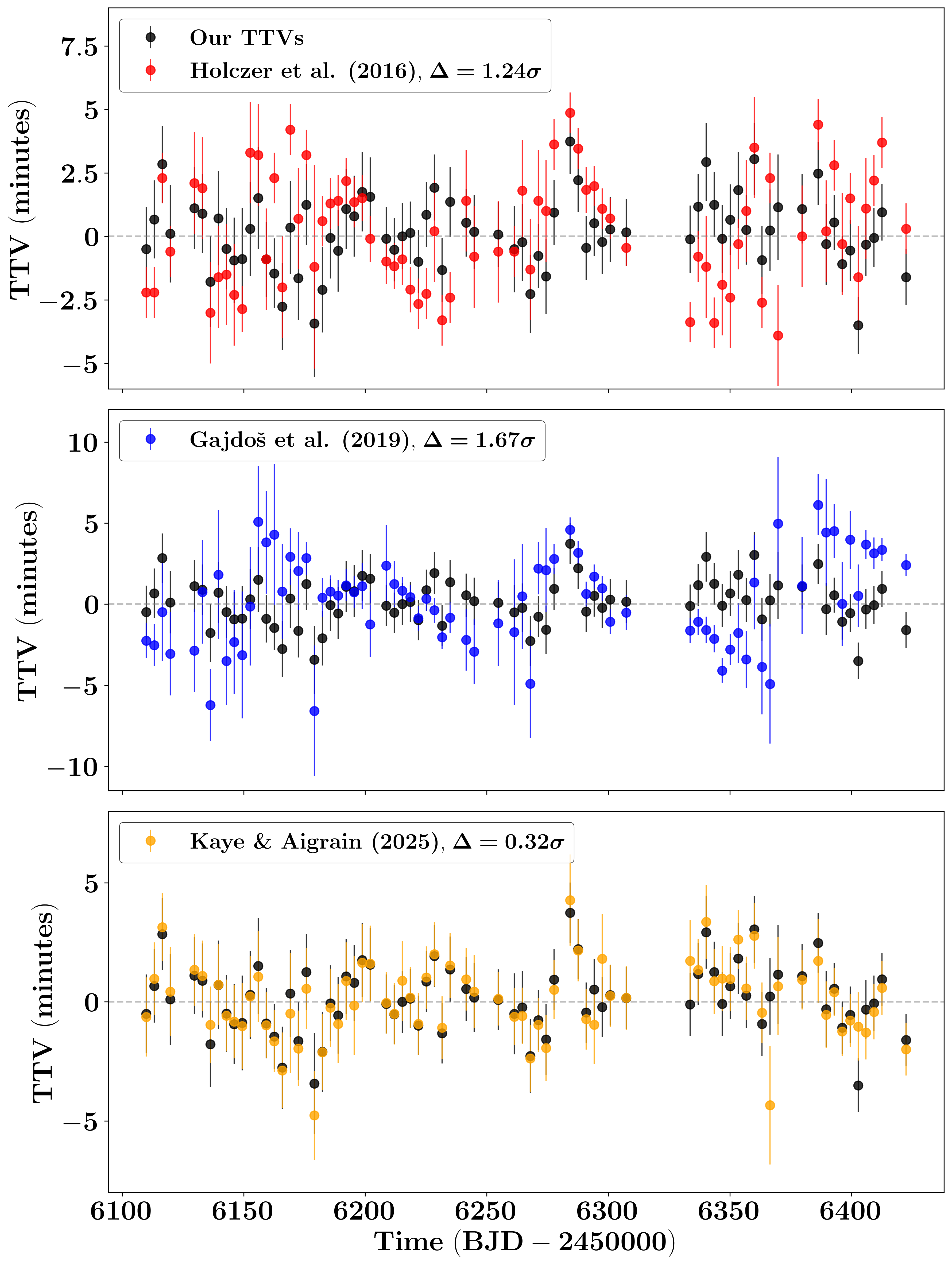}
    \caption{Comparison to published TTVs. Our TTVs are plotted with black circles. TTVs reported by \cite{HolczerMazeh2016}, \cite{GajdosVanko2019}, and \cite{KayeAigrain2025} are plotted in red, blue, and yellow circles, respectively. The $\Delta$ values in the legend represent the average difference between each study's TTVs and ours, normalized by the measurement uncertainties of our TTVs. }
    \label{fig:ttv_study}
\end{figure}

Our reanalysis of the Kepler photometry resulted in a decreased TTV amplitude compared to that reported by \citet{HolczerMazeh2016} and \cite{GajdosVanko2019}. Conversely, \cite{KayeAigrain2025} found an insignificant TTV signal similar to our results based on both Kepler and TESS photometry. Our Kepler TTVs differ from those of \cite{HolczerMazeh2016} by an average of $1.24\sigma$, from \cite{GajdosVanko2019} by an average of $1.67\sigma$, and from \cite{KayeAigrain2025} by an average of $0.32\sigma$ (see Figure \ref{fig:ttv_study}). In this section, we discuss the key differences between the three published studies and our analysis with respect to (1) Kepler pipeline versions, (2) detrending methods, (3) transit midtime fitting method, and (4) TTV significance determination.

First, we note the differences in Kepler pipeline versions used by each study. \cite{HolczerMazeh2016} used an earlier version of the Kepler data, while \cite{GajdosVanko2019}, \cite{KayeAigrain2025}, and our study all used Data Release 25, based on SOC Pipeline 9.3. Since the publication of \cite{HolczerMazeh2016}, the Kepler Pre-search Data Conditioning (PDC) pipeline has been significantly updated. The PDC module in SOC 9.3 implemented a Bayesian fit to high-frequency wavelet bands (Band-3) and developed a method to identify and remove systematic spikes across multiple targets, resulting in noticeable improvements in the quality of the long-cadence data for some quarters \citep{ThompsonCaldwell2016}.

Second, we employed a detrending method using Gaussian kernel convolution as described in \S\ref{Kepler} to better capture long-term systematic trends. In contrast, other studies removed systematic trends by extracting individual transit windows, masking in-transit data, and fitting a polynomial to the out-of-transit data. \cite{HolczerMazeh2016}, \cite{GajdosVanko2019}, \cite{KayeAigrain2025} used first-, second-, and third-degree polynomials, respectively. We confirmed that our detrending method does not significantly affect the transit midtimes by rederiving them using all three forms of polynomial detrending listed above and verifying that the two sets of transit midtimes differ by $<0.35\sigma$ on average in all cases. We repeated our dynamical fits with the updated transit times from the second order polynomial fit along with our new ground-based transit times and confirmed that the one-planet model is still preferred by approximately 2.5$\sigma$ ($\Delta \log \mathcal{Z}= -1.93$).

Third, the method used to derive individual transit timing measurements varies among studies. \citet{HolczerMazeh2016} determined the individual best-fit transit times using a grid search method and then used the inverted Hessian matrix calculated at the best fit location to yield transit timing uncertainties with a median value of 1.0 minutes. However, this optimal estimation approach approximates the global posterior probability distribution as an n-dimensional Gaussian, and therefore may yield inaccurate results for non-Gaussian distributions \citep{SiviaSkilling2006}.  In contrast, \cite{GajdosVanko2019}, \cite{KayeAigrain2025}, and our study all utilized MCMC fits to determine the individual midtime measurements and corresponding uncertainties. We note that the median midtime measurement uncertainties of all three studies are in good agreement: 1.6 minutes \citep{GajdosVanko2019}, 1.4 minutes \citep{KayeAigrain2025}, 1.4 minutes (our study). 

Fourth, \citet{HolczerMazeh2016} allowed both the transit depth and duration to vary for each individual transit event and reported statistically significant fractional transit depth and duration variations. This may indicate that the stroboscopic sampling of these data \citep{SzaboSzabo2013} is biasing their estimates of both the transit shapes and midtimes. In comparison, \cite{GajdosVanko2019},  \cite{KayeAigrain2025}, and our study all fixed the transit shape when solving for the individual transit midtimes. However, when we tested the impact of this assumption by allowing the transit shape to vary when fitting each individual transit event, we found that it had a negligible effect on our derived transit times (average difference of $0.27 \sigma$). 

Finally, each study used a different method for determining the statistical significance of the TTV signal. \citet{HolczerMazeh2016} used a MCMC sampler to fit a periodic sinusoidal function with a linear trend, yielding a TTV amplitude of $2.120 \pm 0.680$ minutes. \citet{GajdosVanko2019} used a sinusoidal model fit by a simple Levenberg-Marquardt algorithm, obtaining a TTV amplitude of $1.9 \pm 0.2$ minutes.  Although they did not specify explicitly, this suggests that they derived the uncertainties on their fitted amplitude using the Hessian matrix (i.e., the optimal estimation method).  As discussed previously, this can lead to inaccurate results for significantly non-Gaussian posterior probability distributions. \cite{KayeAigrain2025} analyzed the O-C residuals with a Lomb-Scargle periodogram and calculated the false alarm probability (FAP) for the peak power, where FAP $<0.05$
indicates a statistically significant TTV signal. Using this method, they determined that Kepler-1624b lacks periodic TTVs (FAP = 0.684). In comparison, we used an N-body integrator to fit our data with both one- and two-planet models and calculated the corresponding Bayes factor between the two fits.

In summary, we are unable to definitively identify the root cause of the decreased TTV amplitude in our study and that of \cite{KayeAigrain2025} relative to previous studies by \cite{HolczerMazeh2016} and \cite{GajdosVanko2019}.  However, the overall good agreement between our conclusions and those of \cite{KayeAigrain2025} suggests that studies using the photometry from the latest version of the Kepler pipeline together with MCMC or other equivalent methods for model fitting should generally reach consistent conclusions about the statistical significance of TTV signals in the Kepler data. Our conclusions for Kepler-1624b are further strengthened by the addition of new transit observations from Palomar, LCOGT, and TESS, which all appear to be well-matched by a single-planet model.

\subsection{Updated transit parameters}
We compare the transit shape parameters derived from our joint fit to the Kepler and Palomar photometry to those reported by \cite{MortonBryson2016}. We find that the $R_p/R_*= 0.10992^{+0.00122}_{-0.00042}$ reported by \cite{MortonBryson2016} is consistent within $1\sigma$ with both our ${R_p/R_*}_{\textrm{\scriptsize WIRC}}$ and ${R_p/R_*}_{\textrm{\scriptsize Kepler}}$ values. The $a/R_* = 15.94 \pm 0.30$ and $b = 0.0358^{+0.196}_{-0.036}$ values reported by \cite{MortonBryson2016} differ by $3.1\sigma$ and $3.8 \sigma$ with our values, respectively. However, this is not surprising given the relatively poor sampling of the transit shape in the Kepler long cadence photometry. We further find that our new $J$ band WIRC photometry allows us to measure $R_p/R_*$ with an uncertainty that is approximately a factor of two smaller than Kepler photometry, demonstrating the importance of high-cadence sampling during ingress and egress for constraining the transit shape parameters. We adopt the $R_p/R_*$ value from the WIRC photometry when calculating $R_p$, as this transit provides the tightest constraint on this parameter. Our new transit photometry extends the baseline of the Kepler photometry by 11 years, resulting in an approximately $20\times$ reduction in the uncertainty on the planet’s orbital period.

\section{Conclusion \label{sec: conclusion}}

In this study, we carried out a detailed reexamination of Kepler-1624b, which appeared to be an extremely rare giant planet orbiting an M dwarf host with a near-resonant companion. We reanalyzed archival Kepler data to retrieve updated transit midtimes for Kepler-1624b and incorporated new TTV measurements that extend the previous TTV baseline by 11 years. We obtained three new ground-based transit observations from Palomar/WIRC, LCOGT/SINISTRO, and LCOGT/MuSCAT3, respectively, and also analyzed recent space-based observations from TESS. We fit these data with a hybrid photodynamical model that simultaneously fits the retrieved midtime measurements from Kepler, Palomar, and LCOGT and directly compares the predicted light curve models to the TESS photometry, which has a lower statistical significance for individual transit events. Using nested sampling, we calculated the Bayes factor between a two-planet model and a one-planet model. Although \citet{HolczerMazeh2016} found that Kepler-1624b had a statistically significant long-term TTV signal suggesting the presence of a nearby planetary companion, our results show that the significance of the TTV signal decreased precipitously in our updated analysis, which prefers the one-planet model by $2.73\sigma$. This is in agreement with the broader study on TTVs by \citet{KayeAigrain2025}, in which 4 out of 11 systems previously known to have TTVs from Kepler or K2 were not determined with statistical significance when including TESS data, including Kepler-1624b. 

Our study underscores the importance of a careful reanalysis of systems with low, few-minute TTV amplitudes identified in TTV catalogs. It is ultimately not surprising that the TTVs prefer the one-planet model, given that our search of archival data indicates that near-resonant companions to gas giants appear to be rare across a broad range of host star mass. 
However, the small sample of gas giants known to orbit M dwarfs limits our ability to rigorously quantify the nearby companion rate for these systems and to determine whether it is consistent with the rate for warm or hot Jupiters orbiting more massive stars. Moving forward, it would be particularly informative to carry out a more rigorous and extensive search for both TTV and radial velocity companions to gas giants orbiting M dwarfs. The architectures of these systems can provide valuable insights into the formation and migration pathways for gas giant planets around low-mass stars.

\section*{acknowledgments}
We are grateful to Yayaati Chachan for helpful conversations about our current understanding of planet formation around low-mass stars. We thank the anonymous referee for reviewing this paper and for providing valuable feedback. We appreciate Pavol Gajdo\v{s}, Laurel Kaye, and Suzanne Aigrain for generously sharing information about their analysis.

This work makes use of observations from the Las Cumbres Observatory global telescope network. This paper includes data collected by the Kepler mission and obtained from the MAST data archive at the Space Telescope Science Institute (STScI). 
Funding for the Kepler mission is provided by the NASA Science Mission Directorate. STScI is operated by the Association of Universities for Research in Astronomy, Inc., under NASA contract NAS 5–26555. This paper includes data collected with the TESS mission, obtained from the MAST data archive at the Space Telescope Science Institute (STScI). 
Funding for the TESS mission is provided by the NASA Explorer Program. STScI is operated by the Association of Universities for Research in Astronomy, Inc., under NASA contract NAS 5–26555.

This research was supported from the Wilf Family Discovery Fund in Space and Planetary Science established by the Wilf Family Foundation.  This research has made use of the NASA Exoplanet Archive and the Exoplanet Follow-up Observation Program website, which are operated by the California Institute of Technology, under contract with the National Aeronautics and Space Administration under the Exoplanet Exploration Program.

This publication makes use of data products from the Two Micron All Sky Survey, which is a joint project of the University of Massachusetts and the Infrared Processing and Analysis Center/California Institute of Technology, funded by the National Aeronautics and Space Administration and the National Science Foundation.

This research has made use of the Exoplanet Follow-up Observation Program (ExoFOP; DOI: 10.26134/ExoFOP5) website, which is operated by the California Institute of Technology, under contract with the National Aeronautics and Space Administration under the Exoplanet Exploration Program.

This work makes use of observations from the LCOGT network. Part of the LCOGT telescope time was granted by NOIRLab through the Mid-Scale Innovations Program (MSIP). MSIP is funded by NSF.

This paper is based on observations made with the MuSCAT3 instrument, developed by the Astrobiology Center and under financial supports by JSPS KAKENHI (JP18H05439) and JST PRESTO (JPMJPR1775), at Faulkes Telescope North on Maui, HI, operated by the Las Cumbres Observatory.

This work is partly supported by JSPS KAKENHI Grant Numbers JP24H00017, JP24K00689, and JSPS Bilateral Program Number
JPJSBP120249910.

This material is based upon work supported by the National Science Foundation Graduate Research Fellowship Program under Grant No.~DGE‐1745301.

\bibliography{sample631}{}
\bibliographystyle{aasjournal}

\end{document}